# The Adoption of ICT Powered Healthcare Technologies towards Managing Global Pandemics

Navod Neranjan Thilakarathne, Mohan Krishna Kagita, Thippa Reddy Gadekallu, Praveen Kumar Reddy Maddikunta

*Department of ICT, University of Colombo, Sri Lanka*

*School of Computing and Mathematics, Charles Sturt University, Melbourne*

*School of Information Technology and Engg, Vellore Institute of Technology, India*

---

**Abstract**

Pandemic is an outbreak that happens over a large geographic area affecting a greater portion of the population as new pathogens appear for which people have less immune and no vaccines are available. It can spread from person to person in a very short time, and in fact, the health workers are at greater risk of infection because of the patients who carry the disease. In the 21st century, where everyone is connected through digital technologies, Information and Communication Technology (ICT) plays a critical role in improving health care for individuals and larger communities. ICT has currently been severed in a variety of application domains which signifies its importance as a major technological paradigm, and it has drawn higher attention for its potential to alleviate the burden on healthcare systems caused by a rise in chronic diseases, aging and increased population and pandemic situations. This paper surveys and offers substantial knowledge about how effective ICT Healthcare strategy can be used to manage global pandemics by presenting a four-phased framework, which can be deployed to alleviate the strain on healthcare during a pandemic. In addition, we discuss how ICT powered technologies can be used towards managing a pandemic during the transformation of simple disease outbreak into a global pandemic.

*Keywords:* ICT, Global Pandemic, Public Health, eHealth, Information and Communication Technology

---

# 1. INTRODUCTION

Over the years, there has been a huge doubt among scientists and medical researchers as to the exact definition of a pandemic, but one thing that everyone agrees is that it describes the widespread occurrence of a disease that spawns large geographical areas, which people have less resistance and no remedies available for ailment [3]. For example, cholera, bionic plague, smallpox, and influenza are some of the most devastating disease outbreaks that has occurred in human history [19]. The impact of a global pandemic is far beyond the scope of public health and medical care due to the heavy burden it will pose on the healthcare resources. It brings mega-scale economic losses and causes social inconsistency and instability[35].

There is no question that globalization in the 21st century has changed our relationships with the biological world radically. As a result, a novel pathogenic infection discovered in one part of the globe can be easily carried thousands of miles away in a single day. With the continuous improvement of the living standards of people in the last decade, the occurrence and development of pandemic events have posed worldwide problems and attracted greater attention as a global threat, even more than terrorism. Examples include the SARS virus outbreak in Hong Kong in 2003, the global pandemic H1N1 flu in 2009 [18]. Advances in global air travel, agricultural technology, urbanization, and mega-scale emissions all make it possible for infectious diseases to emerge and spread [23]. Around the same time, ICT plays a major role in detecting, mapping, understanding, handling, treating, and perceiving global pandemics. Digital communication technologies are playing an increasingly important role in various aspects of global pandemic surveillance, response, introducing novel opportunities to mitigate risks and enhance the efficiency of response [3][6].

Especially during the recent COVID-19 virus outbreak in 2019 and the H1N1 virus outbreak in 2009, it is no doubt that it posed a huge burden on healthcare and has overwhelmed the healthcare infrastructure at the regional level up to the international level. Subsequently, a remedy is required to relieve the strain on healthcare systems whilst continuing to provide high-quality care for contagious patients. With the aid of ICT and related technology, the majority of the issues associated with aging populations, the rise of chronic illnesses, the lack of medical staff can be solved with a minimum effort. Modern measurement devices for medical treatment, such as blood pressure, blood glucose level, heart rate, body temperature, and vari-ous wearable incorporate communication capabilities[39]. They can join and communicate with the remote networks wireless so that patients can be monitored remotely, or medical staff is aware of the patient's condition regardless of the patient's location[42]. As such, ICT will play a key role in strengthening health care for individuals and communities around the world[44]. By offering new and more effective ways to view, communicate, and store information, ICT-powered technologies will help to minimize gaps between healthcare and patients, thus ensuring better quality care in less time. ICT also offers the capacity to boost the quality of the health care systems and eliminate medical errors [26].

The main objective of this paper is to study how effective ICT healthcare strategy can be used to managing global pandemic events, by relieving the strain on healthcare. For the time being, diverse distributed healthcare systems gather, interpret, and interact with each other in real-time, allowing vast volumes of data to be collected, processed, and analyzed effectively. This dependency of healthcare on ICT is increasing day by day, thus it guarantees better patient care, relives the strain of healthcare resources, and provide better analytical within less time for less cost.

This paper is organized into four sections. Following this introduction, a second section describes the literature related to the effective use of ICT in healthcare. The third section presents our proposed ICT powered strategic framework for managing global pandemics and describes each of its components in detail including how to adopt ICT healthcare technologies during the transformation of disease outbreak into a pandemic. In the fourth section, we present the limitations of our study and finally we conclude our study with the conclusion and key findings of the study.

## 2. LITERATURE REVIEW

During a pandemic, it is really hard to predict the anticipated needs of healthcare when the spawning of the disease outbreak is high and if the origin source of the disease outbreak is not properly identified nor the dissecting the outbreak cannot do timely. The traditional healthcare services and applications will risk healthcare workers and put more strain on healthcare facilities and resources hence novel innovative technologies like remote patient monitoring, cloud-based electronic health record systems needed to be employed[40]. Countries with poor sanitary conditions and limited or scarce resources to encounter and defeat these diseases are particularly likely to be severely affected by them (e.g. - Ebola Outbreak in West Africa during 2014-2016). Besides that, due to globalization, the increasingly growing international travel and trade of goods speeds up the spread of pandemics [25].

In order to understand about a pandemic and how it can be successfully managed, firstly the occurrence of a pandemic needed to be described. Based on the sources form the World Health Organization (WHO), they have stated that the evolution of a pandemic can be dissected into three phases as pre pandemic, emergence of a pandemic and the declaration as a pandemic where it can be considered as a global threat [27].Based on these phases, different efforts need to put forward towards managing a pandemic. Over the years, ICT has evolved to integrate with various analytical tools and social media platforms like Facebook for extracting information from such platforms to support the prevention of disease spreading using social media analytics and it involves the use of Geographical Information Systems (GIS) to help in epidemiology. In [1], author has presented a brief discussion about how ICT can be used to detect, surveillance, and prevention of novel zoonotic disease outbreaks at the global and national level. This author has further discussed how open-source and commercial technology can be used in the work of pandemic prevention and protection of human health. It is no doubt that prevention is always better than letting some bad thing to happen. In order to successfully prevent a disease outbreak before it becomes a global pandemic, a nation must have an effective public health surveillance system (PHS) placed. In [14], authors have discussed the importance of ICT powered public health surveillance in order to face emerging infectious disease threats, evolving environmental and behavioral risks, and ever-changing epidemiologic trends in their study. In [23], authors have discussed one aspect of ICT powered eHealth that is public health in related to pandemic control. They presented a five-layered framework to address the management issues of eHealth based on the example of a pandemic control system. An eHealth is an application of ICT that can reduce the effects of a pandemic by enhancing pandemic surveillance and monitoring and improving the efficiency of medical procedures such as efficient reporting using electronic health records systems (EHR). In [17], authors have developed a novel method for eHealth readiness evaluation of a pandemic from the perspective of healthcare organizations and providers on their study.

In [8], authors have discussed ICT-driven solutions for local and global communities aimed at facilitating rapid response to emergencies in public health. In terms of pandemic outbreaks like HIV / AIDS, they have discussed the use of ICT to meet the training and educational needs of public and health workers. They also discussed how mobile devices such as pagers, cell phones, personal digital assistants, tablet computers can play a key role in emergencies, which signifies the use of ICT. Using such devices in healthcare is known as mobile health (mHelath). They have further mentioned that as per the following reasons, these mHelath related technologies are highly effective, when managing pandemics.

1. Mobile devices are reachable anywhere and at any time.
2. Mobile devices can be traceable through GPS.
3. Mobile devices can quickly obtain information (photos, video footages) and communicate under any situation.

In [22], authors have conducted a research on China's emergency response to the SARS crisis, stating that it is necessary to protect global public health and it should be coordinated through global collaboration. Also, they have noted that developing countries possess comparatively fewer resources to cope with emergencies in public health, such as fewer ICU facilities and medical staff. In [9], authors have discussed the importance of distributed digital healthcare manufacturing technologies in their analysis to plan for the next pandemic. This study reviews the literature required in a pandemic like COVID 19 to build open hardware designs. It tests the readiness of the top twenty technologies requested by the Government of India as an example and results show that most of the actual medical devices have had some production of open source, but only 15 percent of the supporting technologies that make the open-source technology possible are publicly available. Their findings indicate that substantial work is still required to provide open source pathways for the development of all appropriate medical hardware during pandemics. In [12], authors have suggested an architecture for the implementation of a cloud-based healthcare support system, as the cloud evolves as an innovative solution with the advantages of improved service quality, reduced costs, and flexibility. In [13], authors have demonstrated how an SMS-based mobile application can be integrated with eHealth, enabling influenza pandemic surveillance in developing countries as a possible eHealth facility for the identification and monitoring of possible pandemic strains

In [15], authors have discussed the development of a compact, user-friendly, and cost-effective point-of-care (POC) diagnostic system. They demonstrated a simple, user-friendly, and inexpensive Internet of Things (IoT) enabled system based on a miniaturized polymerase chain reaction device. The resulting data generated from the device is automatically transferred to an android based smartphone via a Bluetooth interface and then transmitted wirelessly to a global network, making the test results immediately accessible anywhere in the world. In [5], authors have discussed about using an urban intelligence as a resource that analyses knowledge at the city level using data science approaches and discusses its role in the response to pandemics. In [23], authors have reviewed the empirical uses of communication technology in humanitarian and pandemic response, and in particular during the 2014 Ebola response, to suggest a three-part conceptual model for the new informatics of pandemic response.

## 3. PROPOSED STRATEGIC FRAMEWORK FOR MANAGING GLOBAL PANDEMICS

Pandemics may have many catastrophic implications, such as loss of productive workforce and social chaos. Countries with poor sanitary conditions and limited or scarce resources for encountering and defeating such diseases can be severely affected, if not managed well [25]. Moreover, due to the rapidly growing international travel and trade of goods speeds up the spread of a disease outbreak, making it very easy to turn them into pandemics within a short time. This requires tight international cooperation and extending national response capacities to effectively manage pandemics.

The Internet and related technologies provide a new medium for the dissemination of information and provide new ways for institutions, health professionals, and health care providers to interact and collaborate [1]. ICT plays a prominent role in humanitarian operations, and specifically, these groundbreaking solutions are convincing and have attracted much attention worldwide in response to international pandemics. After having introduced how ICT based technologies and related disruptive innovations in healthcare involve in pandemic management through the state of the art review, we will now present our strategic framework for effective pandemic management. We have identified that the following key components needed to be included and initiated in an effective ICT strategy and it should be supported and empowered by national and international level stakeholders such as government authorities , international authorities and health service providers as well as the general public for withstanding against global pandemics as shown in Figure 1.

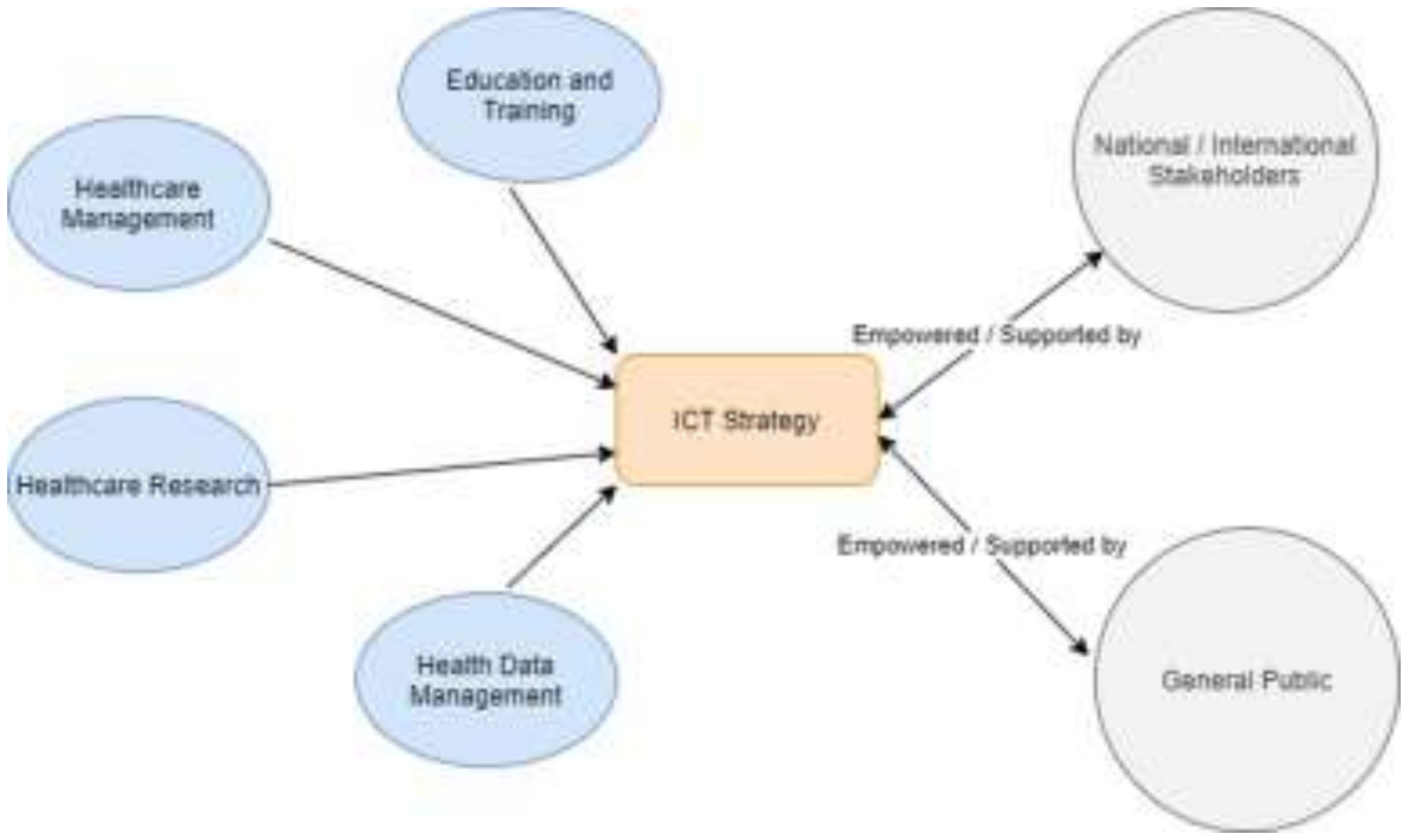

Figure 1: Four phased ICT Strategy.

Based on the Figure 1 we can dissect and discuss the four key components of proposed ICT strategic framework as follows.

### *3.1. Education and training*

Owing to the rapid advancement of technology such as Web 2.0 and beyond, cloud computing, smart mobile devices, people can easily seek, connect, learn interact with others in a very short time in this globalized digital age[30]. This ensures the accessibility and availability of education to all without frontiers. Health education creates awareness among the general public regarding communicable diseases, health status, prevention measures, ailments, and various current diagnostic procedures in the context of health care.

By having a thorough knowledge about deadly disease outbreak before it spawns, the general public can take precautions themselves such as wearing masks and keep social distancing to protect from influenza pandemics. On the other hand, healthcare workers are aware of the precautionary measures that they should take since they are the ones who interact with at-risk patients most. In addition, mHelath linked mobile applications, augmented reality-based simulations can be used to train health workers in the field and the general public, so people are well prepared and trained in advance, and they are well aware of how to survive during the pandemic.

### *3.2. Healthcare management*

ICT allows the hospital management and authorities to successfully direct the hospitals under any circumstances. It helps authorities to resolve the difficulties that hospitals face, during a pandemic as discussed on the following.

1. Rapid response from health-care organizations and officials is expected during the time of a pandemic and it involves their active participation in pandemic surveillance and medical practice. This surveillance and control include case reporting to the local health department or authorities, obtaining and supplying patient details for epidemiological investigations, collecting any health alerts provided by health authorities, and disseminating this information to the general public [1,12].

2. Medical capacity and resources would be considerably challenged during the time of a pandemic. Most challenges are based on the efficiency of documentation (e.g., retrieval, updating, and storage of clinical data). Health care providers will experience a rise in workload during a pandemic and there will be increased demands on medications and prescriptions for community members. The use of electronic health recording systems (EHR), cloud-based electronic health record systems will contribute to efficient monitoring and sharing of health records for patients. In an environment where healthcare services are offered at multiple locations by a variety of healthcare professionals, patient health records must be accessible to all of them, thus deploying EHR systems would be highly beneficial [1,13].

3. The situation will get more worsen when a deadly contagious Influenza virus spawns quickly by affecting a large portion of populations, where a significant amount of medical resources and workers are needed[53]. If there are no adequate ICU facilities, medical practitioners compared to the number of patients, healthcare facilities will be significantly challenged. In such kinds of scenarios, it is better to go for remote patient monitoring and diagnostic and doing more point-of-care diagnostic tests where we can adopt more ICT related technologies to relieve the strain on healthcare resources.

Ultimately, ICT helps management and authorities to improve patient safety and satisfaction, adapt to the latest technologies, and gain better insight about health statistics in the community under a pandemic situation.

### 3.3. Healthcare research

In healthcare research, ICT helps to define potential preventive steps to eliminate and reduce future disease spread. This also saves many people's lives by providing care in advance. Incorporating methods like data science and various methodologies for epidemic simulation, clinical decision support, and earlier identification of diseases help to identify any patterns, forecasting about the outbreak, and providing tailored treatment plans to patients during a pandemic[33]. The conventional manual healthcare processes can be eliminated through ICT-based research, and new models can be created for successful quality care.

### 3.4. Health data management

For hospitals, the basic application of ICT is to store medical data electronically. It conveniently helps to get the details back[4]. The data can be transmitted to the doctors for consultation using relevant ICT technologies. The patient can have in-hand medical records which can be used anywhere, at any time. Medical ability and resources will be considerably challenged during a pandemic, as stated earlier and difficulties lie in the quality of reporting, the distribution of relevant information to the parties concerned, the correct analysis of patient data. Therefore, adoption of EHR systems, cloud-based data management systems, cloud-based data analytics will create huge advantages where it will cut down the manual work and time, increase efficiency and effectiveness when facing pandemic situations[41].

Moreover, based on the transformation of a simple disease outbreak to a global pandemic, a pandemic should pass several phases chronographically. That is prevention, detection and verification, response and finally the continuous sustained response phase[32]. In each phase, ICT technologies can be utilized in several ways depending on the context for effective results. Following Figure 2 describes how ICT related technologies can be adopted for managing a pandemic chronographically[49].

As depicted in the Figure 2 based on each phase, the actions we need to take will be vary based on the context for better results. In the next section, we discuss about this each of the actions and how ICT powered technologies can be used for better results.

### 3.5. Public health surveillance

Public health surveillance is fundamental to the disease prevention process and is the process of continuous, systematic collection, review, and eval-uation of health data to identify and track a health incident. Surveillance data are used both to evaluate the need for intervention in the field of public health and to assess program effectiveness. Public health monitoring is known as an early warning system, a rudimentary indicator of irregular forms of disease emerging [13]. During the time of the pandemic, rapid response from healthcare organizations and authorities is required. It calls for their active participation in pandemic monitoring and medical practice. Such monitoring and control require case reporting to local health departments or authorities, accessing and providing patient information for epidemiological investigations, capturing any health warnings issued by health authorities, and disseminating this information to the general public [1][12]. As an example of this public health surveillance drones have been used during COVID-19 pandemic, to ensure quarantine compliance and to monitor whether people are equipped with protective gear such as face masks[54]. Further incorporating web-based real-time surveillance will help to real-time monitor of disease activities. (e.g., Use of Google Trends, Google Flu Trends to identify seasonal influenza activity)

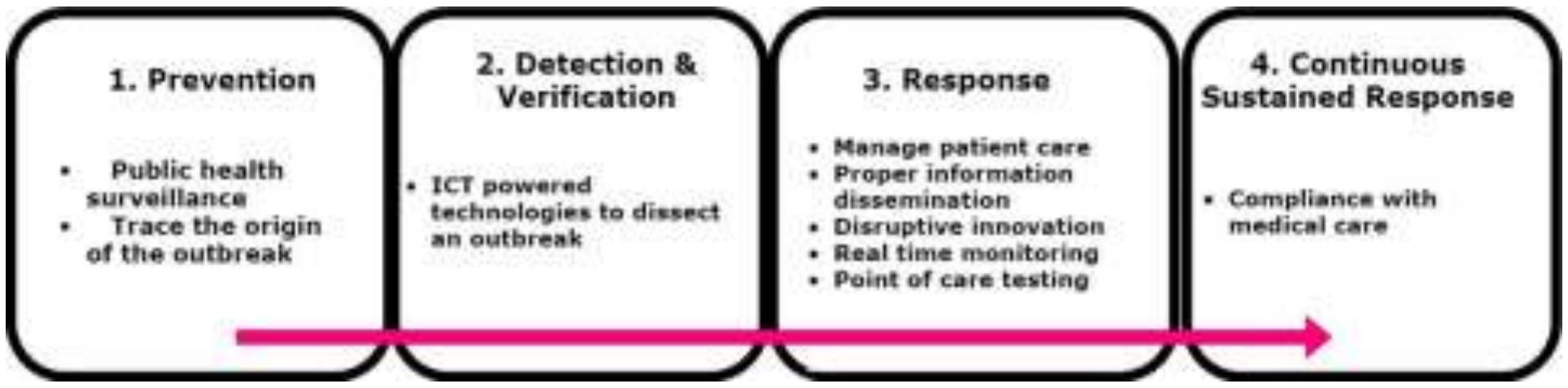

Figure 2: Employment of Context based ICT Technologies during a Pandemic.

### *3.6. Trace the origin of the outbreak*

ICT can be used during a pandemic to track the sources of an outbreak. A recent study by MIT researchers utilized aggregated mobile phone data to trace the spread of dengue virus in Singapore during 2013 and 2014 in granular details of short distances and periods. In these cases, it can assist epidemiologists in their search for patient zero and it can help to classify all the individuals that have come into contact with the infected patients and thus also be infected [28].

Using ICT powered technologies to dissect an outbreak With the various and complex datasets gathered about the disease outbreak, we can effectively analyze the outbreak using data science methodologies[50]. Apart from that following methodologies can be incorporated whilst dissecting an outbreak. 1. Epidemic simulation assessment of disease spread determinants 2. Participatory epidemiology using social media

### *3.7. Using ICT to manage patient care*

Additionally, the scalability of ICT is useful for monitoring all patients at high risk to warrant quarantine but not bad enough to warrant in-hospital treatment, particularly during an influenza pandemic. With ICT-based technology adoption, patients can track their temperature and blood pressure using wearable devices and upload data to the cloud for review via their mobile devices. This not only helps healthcare staff to gather more data with less time but also reduces the risk of cross-infection with patients, implying the benefits of ICT usage[51]. On the other hand, daily, check-ups for hospitalized patients can be done remotely and it will enable diagnosing patients using remote point of care testing facilities[48].

### *3.8. Proper information dissemination and connectivity*

Clear communications to the general public during a pandemic are important to ensure that they are aware and advised of taking necessary measures, and that accurate information is collected. Nowadays, most telecommunications devices are user friendly and used by a huge worldwide population that has reduced the contact gap to a zero point. Accessibility to information has now become easier with the use of ICT, and people are therefore more comfortable when accessing healthcare[52]. Push notifications based on Smart Phones, SMS alerts are able to use the available technology to easily hit the population with updated information and, most importantly, to refresh it rapidly as the hazards shift. With the use of these resources, it is clear that we are able to reach a desperate population and make them aware of the challenges and therapies for some of these diseases. It is important to note that while technology and cell phones are becoming more prevalent worldwide the usage of smartphones in rural areas is still limited, thus adopting an SMS based information dissemination mechanism will be highly useful [24].

### 3.9. Disruptive innovation

Authorities especially need to encourage and fund the general public for coming up with new innovations by lessening the restrictions with policy-making organizations and government bodies, as some of the government procedures consume a lot of time to process innovations and introduce it to the market. A novel innovations that are being tested include analyzing data from cell phone towers to track users who have been closed to the known case of the virus [24].As the time of writing this, where COVID-19 has overwhelmed the entire globe we can see that there is a clear rise among university students, researchers, innovators for inventing such ICT powered solutions for healthcare. (e.g., IoT based point of care diagnostic systems, remote patient monitoring tools)

### 3.10. Real time monitoring

ICT has also been used in the continuous remote monitoring of in-home patients with medical conditions such as hypertension, asthma, and diabetes[45]. In hospitals, telemetry, the transmission of measurements like heartbeat, temperature, blood pressure, and oxygen saturation level from wearables and various ingestible sensors on patients to the central monitoring has been used to track large numbers of patients with limited staff[36]. With the increasing advancement in communication technologies, many revolutionary forms of healthcare systems with body sensor networks (BSN), Wireless Sensor Networks (WSN) are used for real-time monitoring[29].

### 3.11. Compliance with medical care and constant monitoring

Constant monitoring of an outbreak for any deadly disease is the most important aspect of any healthcare support program[38]. For instance, H1N1 is an infectious virus that affected a large volume of the population upon spread. During the 20th century alone, the spread of the H1N1 virus occurred three times, affecting 500 million people worldwide. WHO has noted that the root cause of the outbreak is the non-existence of primary H1N1 immunity in humans? To minimize its adverse effects, both pharmaceuticals and non-pharmaceutical strategies needed to be employed. Effectively, ICT tools can be used to prevent and control this virus strain [12].

ICT can also be used to ensure patient compliance with the quarantine. Public health workers can keep track of which patients remained in quarantine and which patients infringed the quarantine. The use of GPS-enabled wearables would also help to monitor them and determine who else may be exposed because of the breach.

In short, ICT has the potential to affect any part of the health sector. The applications of ICT in healthcare has many benefits, such as greater access to complete and reliable patient data, which gathers information to improve diagnosis, avoid mistakes, and thereby save valuable response times. This also leads to quality care for patients. ICT helps to streamline healthcare processes where the public health system is in an absolute mess due to numerous administrative deficiencies, especially in developing nations[43]. ICT healthcare technologies can help to streamline these activities, thus dramatically lowering expenses, lowering manual labour, and reducing the number of doctor visits, which ultimately leads to improved control of healthcare facilities during a crisis such as global pandemics.

## 4. LIMITATIONS

It is often considered that services and applications based on ICT are often considered to be costly, too risky, and distract from the primary objective and purpose of health sector programs[31]. However, to some degree these critiques may be true, however, there are wide range of low-cost options are also available that should be adopted by hospital management and authorities. Implementation of an ICT solution and the adoption of the related technologies frequently entails a radical transition in the healthcare workplace. As a result, resistance may occur from the individual as well as organizational levels, thus it requires proper planning and, providing adequate knowledge to the relevant stakeholders.

## 5. CONCLUSION

Today, technology changes the way we live and leads us towards a sophisticated technical world. It is no doubt that the emerging role of ICT powered technologies has created a huge impact on healthcare[34]. Also incorporating ICT technologies helps to increase the quality of care, improves patient safety, protects data, and decreases operational administrative costs[37]. It is difficult to predict the expected needs of healthcare during a pandemic hence pandemic preparedness models should be adopted in advance. Taking conventional healthcare models into account, they can endanger healthcare workers, and therefore strategies that allow social distancing are critical in reducing the risk of exposure to health workers. As such, innovative ICT based approaches like remote patient consultation can provide minimal capability for evaluation from anywhere is very helpful to perform pre-screening and eliminate crowding at the hospitals[47].

This paper provides a comprehensive study about how effective ICT strategy can be used to manage healthcare services and facilities during a global pandemic. Always we need to know that occurrence of the next pandemic is imminent and unpredictable. We can never be certain of when or where the next pandemic will arise. Finally, we hope this research will be useful for healthcare professionals, researchers, academics, students, and anyone who seeks new knowledge in the area of ICT in healthcare